\documentstyle[12pt,fleqn]{article}
\begin{document}

\begin{center}
{\bf Partial Factorization of Wave Function for A Quantum Dissipation System}

{\it C. P. Sun}

{\it Institute of Theoretical Physics, Academia Sinica, Beijing 100080, P.
R. China and Physics Department, the Chinese University of Hong Kong, Hong
Kong, China}

{\it Y. B. Gao, H.F.Dong}

{\it Institute of Theoretical Physics, Northeast Normal University,
Changchun 130024, P. R. China.}

{\it S. R. Zhao}
\end{center}

{\it Department of Applied Mathematics, University of Western Ontario,
London, }

{\it Ontario, Canada, N6A, 5B7}

\begin{center}
\bigskip\ 

{\bf ABSTRACT}
\end{center}

\noindent The microscopic approach quantum dissipation process presented by
Yu and Sun [Phys. Rev., A49(1994)592, A51(1995)1845] is developed to analyze
the wave function structure of dynamic evolution of a typical dissipative
system, a single mode boson soaked in a bath of many bosons. In this paper,
the wave function of total system is explicitly obtained as a product of two
components of the system and the bath in the coherent state representation.
It not only describes the influence of the bath on the variable of the
system through the Brownian motion, but also manifests the back- action of
the system on the bath and the effects of the mutual interaction among the
bosons of the bath. Due to the back-action, the total wave function can only
be partially factorizable even for the Brownian motion can be ignored in
certain senses, such as the cases with weak coupling and large detuning. 
\bigskip

\newpage\ 

\qquad

\noindent  {\bf 1. Introduction: }The typical microscopic treatment for the
model of quantum dissipation is to consider a harmonic oscillator
interacting with a many-oscillator bath (e.g., in [1-6]). This simple and
rather conventional model has been used to completely clarify the relation
between two different approaches for quantum dissipation process frequently
appearing in the literature, i.e., the system plus bath model and the
time-dependent effective Hamiltonian by Kanai and Calderora [7,8], since Yu
and one (Sun) of the authors wrote down the total wave function explicitly
in a form of direct product of the bath component and the system component
[1,2]. In the discussion, because the mixed variables were chosen to
describe the system and the bath, the wave function only manifested the
influence of the bath on the system through the Brownian broadening of the
width of the wave function for the system, but the back-action of the system
on the bath was not discussed. If there indeed exists the back-action of the
system on the bath, it is reasonable to expect that, for the individual
particles constituting the bath, the mutual couplings among them can
indirectly appear in a second order through coupling the system as an
intermediate process.

Another question is the relation between quantum and classical systems. In
many real situations, the classical or macroscopic states can be represented
by coherent states in the quantum optics and the macroscopic quantum
mechanics. Therefore it is significant to study how the system with an
initial coherent state evolves if it really has a macroscopic or classical
meanings and to test if it can move in the classical orbits.

In this paper, the back-action and mutual couplings with a simple model is
studied and manifested in the coherent representation. In the presence of
both the back-action and the indirect mutual coupling, we also consider the
meaning of the wave function of the dissipative system governed by the
effective Hamiltonian, which is also to be determined in this paper. A very
interesting result is that a factorizable evolution represented by a product
of two coherent states is obtained with one component representing the
system, which evolves according to a classical orbit and possess a classical
meaning. \bigskip\ 

\noindent  {\bf 2. Structure of Wave Function: }This paper will consider the
problem with the simple model consisting of a single mode boson and a bath
of many bosons and its Hamiltonian is written as 
\begin{equation}
\label{1}H=\hbar \omega b^{+}b+\sum\limits_j\hbar \omega _ja_j^{+}a_j+\hbar
\sum\limits_j\left[ \xi _jb^{+}a_j+h.c\right] , 
\end{equation}
where $\xi _j=|\xi _j|e^{i\sigma _j}$'s are the complex coupling constants
and $b^{+}$, $b$, $a_j^{+}$ and $a_j$ are the bosonic creation and
annihilation operators for the system and the bath respectively. This model
can be regarded as a rotating wave approximation of the original oscillator
model with the linear coupling $\sum\nolimits_j\xi _jqx_j$ $\sim
\sum\nolimits_j[\xi _jb^{+}a_j+\xi _jb^{+}a_j^{\dagger }+h.c]$ of the system
coordinate $q$ to the bath variables $x_j$.

To obtain the explicit expression for the wave function of the total system
formed by the system plus the bath, we invoke the well-known solutions [9] 
\begin{equation}
\label{2}b(t)=u(t)b(0)+\sum_j\upsilon _j(t)a_j(0), 
\end{equation}

\begin{equation}
\label{3}a_j(t)=e^{-i\omega _jt}a_j(0)+u_j(t)b(0)+\sum_{s(\neq j)}\upsilon
_{js}(t)a_s(0) 
\end{equation}
of the corresponding Heisenberg equation, where the coefficients are 
\begin{equation}
\label{4}u(t)=\exp \left[ -\frac \gamma 2t-i(\omega +\Delta \omega )t\right]
, 
\end{equation}

\begin{equation}
\label{5}\upsilon _j(t)=-\xi _j\exp \left( -i\omega _jt\right) \frac{1-\exp
\left[ i\left( \omega _j-\omega -\Delta \omega \right) t-\gamma t/2\right] }{%
\omega +\Delta \omega -\omega _j-i\gamma /2}, 
\end{equation}
\begin{equation}
\label{6}u_j(t)=-\xi _j^{*}e^{-i\omega _jt}\frac{\exp \left[ i\left( \omega
_j-\omega -\Delta \omega \right) t-\gamma t/2\right] -1}{\omega _j- \omega
-\Delta \omega +i\gamma /2}, 
\end{equation}
\begin{equation}
\label{7}\upsilon _{js}(t)=\frac{-\xi _j^{*}\xi _se^{-i\omega _jt}}{\omega
+\Delta \omega -\omega _s-i\gamma /2}\left\{ \frac{1-\exp \left[ i\left(
\omega _j-\omega -\Delta \omega \right) t-\gamma t/2\right] }{\omega +\Delta
\omega -\omega _j-i\gamma /2}-\Lambda \right\} 
\end{equation}
where 
\begin{equation}
\label{8}\Lambda =\left\{ 
\begin{array}{cc}
\frac{\exp \left[ i\left( \omega _j-\omega _s\right) t\right] }{\omega
_j-\omega _s} & \qquad j\neq s \\ 
t & \qquad j=s 
\end{array}
\right. . 
\end{equation}
Here, Lamb shift $\Delta \omega $ can be absorbed into $\omega $ to get the
renormalized frequency $\tilde \omega =\omega +\Delta \omega $ and the
damping constant $\gamma $ is determined by the coupling $\xi _j$and the
spectrum density $\rho (\omega _j)$ of the bath. If we chose an appropriate $%
\rho (\omega _j)$ similar to the Ohmic one by Caldeira and Leggett [3], the
above solution is exact [1,2], otherwise, it can be obtained by the
Wigner-Weisskopf approach [9].

Now, we present a new method in the coherent state representation to
calculate the evolution of the wave function in the Schr\"odinger picture
from the explicit expressions of the canonical operators in the Heisenberg
picture. It is different from that in our previous works [1,2], but quite
direct and effective. If the initial state of the total system is a direct
product $\left| \Psi (0)\right\rangle =\left| \phi \right\rangle \otimes
\prod\nolimits_j\left| \phi _j\right\rangle $ and $U(t)$ is the evolution
operator of the total system, the wave function $\left| \Psi
(t)\right\rangle =U(t)\left| \Psi (0)\right\rangle $ at time $t$ can be
defined by its coherent state representation 
\begin{equation}
\label{9}\Psi (\lambda ,\{\lambda _j\},t)=\langle \lambda ,\{\lambda
_j\}\mid \Psi (t)\rangle =\langle \Psi (0)|U(t)^{\dagger }|\lambda
,\{\lambda _j\}\rangle ^{*}. 
\end{equation}
Here, we used the overcomplete basis 
\begin{equation}
\label{10}|\lambda ,\{\lambda _j\}\rangle =\left| \lambda \right\rangle
\otimes \prod_j\left| \lambda _j\right\rangle =N(\lambda ,\{\lambda
_j\})\exp \left( \lambda b^{\dagger }(0)+\sum_j\lambda _ja_j^{\dagger
}(0)\right) |0\rangle 
\end{equation}
constructed by the coherent states $\left| \lambda \right\rangle $ and $%
|\lambda _j\rangle $ for the annihilation operators $b(0)$ and $a_j(0)$
respectively. Here, the normalization constant $N(\lambda ,\{\lambda _j\})$ $%
=\exp \left( -\frac 12|\lambda |^2-\sum\nolimits_j\frac 12|\lambda
_j|^2\right) $. Then, we turn to obtain an explicit expressions of $%
U(t)|\lambda ,\{\lambda _j\}\rangle $ by considering the role of the
evolution matrix $U(t)$ in the Heisenberg picture. In fact, since $%
U(t)^{\dagger }O(0)U(t)=O(t)$ and $U(t)|0\rangle =|0\rangle $ for a operator 
$O,$ it is easy to obtain 
$$
U(t)^{\dagger }|\lambda ,\{\lambda _j\}\rangle =N\left( \lambda ,\{\lambda
_j\}\right) \exp \left( \lambda b^{\dagger }(t)+\sum_j\lambda _ja_j^{\dagger
}(t)\right) |0\rangle 
$$
\begin{equation}
\label{11}=N\left( \lambda ,\{\lambda _j\}\right) \exp \left( \alpha
(t)b^{\dagger }(0)+\sum_j\lambda \beta (t)a_j^{\dagger }(0)\right) |0\rangle
=\left| \alpha (t)\right\rangle \otimes \prod_j\left| \beta
_j(t)\right\rangle 
\end{equation}
where 
\begin{equation}
\label{12}\alpha (t)=u(t)^{*}\lambda +\sum_j\lambda _ju_j(t)^{*} 
\end{equation}
\begin{equation}
\label{13}\beta _j(t)=e^{i\omega _jt}\lambda _j+v_j(t)^{*}\lambda
+\sum_{s(\neq j)}v_{s,j}(t)^{*}\lambda _s 
\end{equation}
Finally, we obtain a formal factorized wave function for the total system 
\begin{equation}
\label{14}\Psi \left( \lambda ,\{\lambda _j\},t\right) =\phi \left(
u(t)^{*}\lambda +\sum\limits_ju_j(t)^{*}\lambda _j\right) \otimes
\prod\limits_j\phi _j\left( e^{i\omega _jt}\lambda _j+v_j(t)^{*}\lambda
+\sum_{s(\neq j)}v_{s,j}(t)^{*}\lambda _s\right) . 
\end{equation}
\qquad

\qquad \bigskip\ 

\noindent  {\bf 3. Partial Factorization and Effective Hamiltonian}: The
above wave function (14) is not completely factorizable because of the
entanglements of the variables $\lambda $ and $\lambda _j$, which are
implied by the term, $\sum\nolimits_ju_j(t)^{*}\lambda _j$, of the bath
variables $\lambda _j$ modifying the system variable $\lambda $ and the
term, $v_j(t)^{*}\lambda $, of the system modifying the bath one. This
former represents the bath fluctuation due to the Brownian motion while the
later the back-action of the system on the bath. In fact the term, $%
\sum\nolimits_ju_j(t)^{*}\lambda _j$, is caused by the bath fluctuation
operator $B(t)=\sum_j\upsilon _j(t)a_j(0)$ in the system operator $b(t),$
which has a zero thermal average, but a non-zero correlation 
\begin{equation}
\label{15}<B(t)^{\dagger }B(t^{\prime })>=\sum_j4|\xi _j|^2\frac{%
f_j(t,t^{\prime })}{\gamma ^2+4\left( \omega _j-\omega -\Delta \omega
\right) ^2}\left[ \exp \left( \frac{\hbar \omega _j}{k_BT}\right) - 1\right]
^{-1}. 
\end{equation}
The term, $\sum_{j\neq s}v_{s,j}(t)^{*}\lambda _s$, shows the mutual
interactions among the bosons of the bath through the system.
Mathematically, if the coupling is weak with the small $\xi _j,$ the mutual
interactions are the second order as shown by $v_{s,j}(t)^{*}\propto \xi
_j^{*}\xi _s.$ Notice that the main difference between the present result
and that in refs. [1,2] is the back-action and the mutual interactions.

When the fluctuation can be ignored for certain cases, e.g., $\lambda $ is
very large in the initial state and the coupling is weak enough, the
entanglement disappears so that the wave function becomes a product 
\begin{equation}
\label{16}\Psi \left( \lambda ,\{\lambda _j\},t\right) \approx \phi \left(
u(t)^{*}\lambda \right) \prod\limits_{j=1}^N\phi _j\left( e^{i\omega
_jt}\lambda _j+v_j(t)^{*}\lambda \right) . 
\end{equation}
In this case, all the influences of the bath on the system are represented
by the damping constant $\gamma $ and then the wave function is partially
factorizable due to the term $v_j(t)^{*}\lambda $. It is not difficult to
prove that the system component $\phi (u(t)^{*}\lambda )$ is governed by an
effective Hamiltonian, which is also equivalent to the Calderora-Kanai
Hamiltonian.

To prove it, we need to return into the Heisenberg picture by dropping of
the bath operators $a_j(0)$ in $b(t)$, namely, $b(t)$ is replaced by $\tilde
b(t)$ $=u(t)b(0)$. However, $\tilde b(t)^{+}$ and $\tilde b(t)$ are not the
canonical operators since $[\tilde b(t),\tilde b(t)^{+}]=e^{-\gamma t}$.
However, the Bogoliubov transformation gives the general canonical operators 
\begin{equation}
\label{17}A(t)=\alpha \tilde b(t)+\beta \tilde b(t)^{+} 
\end{equation}
satisfying $[A(t),A(t)^{+}]=1$, where 
\begin{equation}
\label{18}\left| \alpha \right| ^2-\left| \beta \right| ^2=\exp \left(
\gamma t\right) . 
\end{equation}
To given the correct Heisenberg equations for operators $A(t)$ and $A(t)^{+}$%
, the effective Hamiltonian is determined by the definition (17) as
time-dependent 
$$
H_{eff}=i\hbar \exp \left( -\gamma t\right) \left[ \left( \bar \alpha \alpha
^{*}-\bar \beta \beta ^{*}\right) A(t)^{+}A(t)+\frac 12\left( \bar \beta
\alpha -\bar \alpha \beta \right) A(t)^{+}A(t)^{+}\right. 
$$
\begin{equation}
\label{19}+\left. \frac 12\left( \bar \alpha ^{*}\beta ^{*}-\bar \beta
^{*}\alpha ^{*}\right) A(t)A(t)\right] 
\end{equation}
where $\bar \alpha =\dot \alpha -(\gamma /2+i\tilde \omega )\alpha $ and $%
\bar \beta =\dot \beta -(\gamma /2-i\tilde \omega )\beta $. Notice that the
number $\bar \alpha \alpha ^{*}-\bar \beta \beta ^{*}=\delta $ should be a
pure imaginary number, i.e. $\delta ^{*}=-\delta $. It is not exotic that
the effective Hamiltonian is not unique because there is only one constraint
(18). Its different forms correspond to different realizations of the
canonical variables.

For instance, a specific solution of Eq. (18), $\alpha $ $=\exp (\gamma
/2+i\varphi )t,$ $\beta =0$, $\varphi =\tilde \omega -\sqrt{\gamma
^2/4+\tilde \omega ^2}$, gives $H_{eff}=\hbar \exp (\gamma t)\Omega
A(t)^{+}A(t)$ with $\Omega =\sqrt{\gamma ^2/4+\tilde \omega ^2}$. By
formally introducing the canonical coordinate $Q=\sqrt{\hbar /2M\Omega }%
\left[ A(t)+A(t)^{+}\right] $ and momentum $P=-i\sqrt{M\Omega \hbar /2}%
\left[ A(t)-A(t)^{+}\right] $ with the varying mass $M=m\exp (\gamma t)$,
this special effective Hamiltonian is just of the form by Calderora and
Kanai. This result can be also given in purely quantized version in
Schr\"odinger picture by the direct calculation of matrix elements, $%
\left\langle \alpha \right| H_{eff}\left| \beta \right\rangle =\left\langle
\alpha \right| i\hbar \left( \partial U(t)/\partial t\right) U(t)^{+}\left|
\beta \right\rangle $. \bigskip\ 

\noindent  {\bf 4. Motion of The Center of Wave Packet with Quantum
Fluctuation: }Let us consider the physical significance of the above wave
function and its entanglements in details. In the representation of
coordinate-momentum, 
\begin{equation}
\label{20}q=\sqrt{\frac \hbar {2\omega }}\left( b+b^{\dagger }\right)
,\qquad p=-i\sqrt{\frac{\hbar \omega }2}\left( b-b^{\dagger }\right) , 
\end{equation}
\begin{equation}
\label{21}x_j=\sqrt{\frac \hbar {2\omega _j}}\left( a_j+a_j^{\dagger
}\right) ,\qquad p_j=-i\sqrt{\frac{\hbar \omega _j}2}\left( a_j-
a_j^{\dagger }\right) , 
\end{equation}
coherent states $\left| \lambda \right\rangle $ and $\left| \lambda
_j\right\rangle $ are understood as the Gaussians of widths $\sqrt{\hbar
/2\omega }$ centered in $q_0=\sqrt{\hbar /2\omega }\left( \lambda +\lambda
^{*}\right) $ and $x_{j0}=\sqrt{\hbar /2\omega }\left( \lambda _j+\lambda
_j^{*}\right) $ respectively. If the initial state of the total system is a
direct product of such Gaussians, 
\begin{equation}
\label{22}\left| \Psi (0)\right\rangle =\left| \lambda =\sqrt{\frac \omega
{2\hbar }}q_0\right\rangle \otimes \prod\limits_j\left| \lambda _j=\sqrt{%
\frac{\omega _j}{2\hbar }}x_{j0}\right\rangle , 
\end{equation}
the wave function at time $t$ 
$$
\left| \Psi (t)\right\rangle =\left| \lambda (t)\right\rangle \otimes
\prod_j\left| \lambda _j(t)\right\rangle =\left| u(t)\sqrt{\frac \omega
{2\hbar }}q_0+\sum_ju_j(t)\sqrt{\frac{\omega _j}{2\hbar }}%
x_{j0}\right\rangle 
$$
\begin{equation}
\label{23}\otimes \prod\limits_j\left| \sqrt{\frac{\omega _j}{2\hbar }}%
x_{j0}e^{i\omega _jt}+v_j(t)\sqrt{\frac \omega {2\hbar }}q_0+\sum_{j(\neq
s)}v_{s,j}(t)\sqrt{\frac{\omega _s}{2\hbar }}x_{s0}\right\rangle 
\end{equation}
defines the position evolution of center of the Gaussian wave packet 
$$
q_c(t)=\sqrt{\frac \hbar {2\omega }}\left( \lambda (t)+\lambda
^{*}(t)\right) 
$$
\begin{equation}
\label{24}=q_0\exp \left( -\frac 12\gamma t\right) \cos \left( \tilde \omega
\right) t+\sum_j|\xi _j|\sqrt{\frac{\omega _j}\omega }\frac{x_{j0}\Theta
_j(t)}{\gamma ^2/4+\left( \omega _j-\tilde \omega \right) ^2}, 
\end{equation}
$$
\Theta _j(t)=\exp \left( -\frac 12\gamma t\right) \left[ \frac \gamma 2\sin
\left( \tilde \omega +\sigma _j\right) t+\left( \omega _j-\tilde \omega
\right) \cos \left( \tilde \omega +\sigma _j\right) t\right] 
$$
\begin{equation}
\label{25}-\left[ \frac \gamma 2\sin \left( \omega _j+\sigma _j\right)
t+\left( \omega _j-\tilde \omega \right) \cos \left( \omega _j+\sigma
_j\right) t\right] . 
\end{equation}

It is known from Eq. (16) that the center of the wave packet moves along the
classical trajectory of a damping harmonic oscillator, which is described by
first term in Eq. (16) and perturbed by the initial displacements $x_{j0}$
of the bath oscillators shown in the second term in Eq. (16). This
fluctuation effect is just the explicit manifestation of the Brownian
motion. For a very large initial displacement, $q_0,$ the weak coupling with
small $\xi _j$ or a sharp spectral distribution of the bath with a large
detuning from the renormalized frequency $\tilde \omega ,$ one can ignore
this fluctuation effect.

Finally, we consider the back-action of the system on the bath and the
mutual interaction among the bosons of the bath through the motion law of
the center of Gaussian for each boson of the bath 
$$
x_{jc}(t)=\sqrt{\frac \hbar {2\omega _j}}\left( \lambda _j(t)+\lambda
_j^{*}(t)\right) 
$$
\begin{equation}
\label{26}=x_{j0}\cos \omega _jt+q_0\sqrt{\frac \omega {\omega _j}}Re
\left( v_j(t)\right) +\sum_{s(\neq j)}Re\left( v_{s,j}(t)\right) 
\sqrt{\frac{\omega _s}{\omega _j}}x_{s0}. 
\end{equation}
Notice that the back-action $q_0\sqrt{\omega /\omega _j}$Re$(v_j(t))$ is
proportional to both the initial displacement $q_0$ times a Lorentz function 
$\propto \left[ \gamma ^2/4+(\omega _j-\tilde \omega )^2\right] ^{- 1}$.
Thus it can not be neglected for large $q_0$ or in the quasi-resonance case
that the bath spectral distribution $\rho (\omega _j)$ peaked in the
renormalized frequency $\tilde \omega $ of the system. The last term $%
\sum_{s(\neq j)}$Re$(v_{s,j}(t))\sqrt{\omega _s/\omega _j}x_{s0}$ is of
second order and explicitly reflects the mutual coupling among the bosons of
the bath, which can be neglected in the first order approximation. \bigskip\ 

\noindent  5. {\bf Summary and Discussions: }In summary, we first mention
that the Langevin approach is an standard treatment for quantum dissipation
process in the present model Hamiltonian [4], but it hardly concerns the
structure of wave function that is essentially important in zero
temperature; the Markoff approximation is also a quite effective density
operator method, but it only considers very a few intuitional picture based
on the classical correspondence and its dissipation-fluctuation relation
directly. This paper take both two aspects of this problem into account and
thus gives a direct and clarified picture for quantum dissipation process
with the most simple model. Our discussion not only concerns the necessary
details in the dynamics of quantum dissipation, but also reveals the roles
of the back-action of bath and the mutual coupling among the bosons of bath.
Notice that the Langevin approach is based on a stochastic equation

\begin{equation}
\label{27}\dot b(t)=\left[ -\frac \gamma 2-i\left( \omega _a+\Delta \omega
\right) \right] b(t)+F(t) 
\end{equation}
with the stochastic force $F(t)=-i\sum_j\xi _ja_j(0)e^{-i\omega _jt}.$ Its
explicit solution is the starting point finding the partially factorizable
wave function of the total system in this paper. In this sense, our study is
a generalization of quantum Langevin theory. As for the Markoff theory, our
explicit solution somewhat can be regarded as the zero-temperature result of
the density matrix approach, but it deal with the dissipation - fluctuation
relation in the framework of wave function. It is straightforward to extend
present discussion to the finite temperature case. Another emphasis of this
paper is the partially factorizable structure of the evolution function,
which clarifies the new meaning of the wave function governed by the
time-dependent effective Hamiltonian by Calderora and Kanai in the presence
of the back-action and the indirect mutual interactions. Notice that this
meaning in our previous works [1, 2] was only understood in the sense of the
complete factorization of wave function.

To conclude this paper we point out that the method used both in this paper
and the previous works [1,2] is very limited to a linear system we
considered before, such as the harmonic oscillator, the inverse harmonic
oscillator (the harmonic oscillator with image frequency, $\omega
\rightarrow i\omega )$ and a linear potential for a constant force. This is
because, only for these systems, the solutions of canonical Heisenberg
operators are the linear combinations of the system variable and the bath
variables and thus the wave functions are factorizable in the appropriate
representations. There still exists the difficulties in principle to
generalize the idea and method of this paper to the nonlinear cases except
those systems that can be reasonably linearized.

{\it One (CPS) of the authors wish to express his sincere thanks to K. Young
for inviting him to visit Hong Kong as a C. N. Yang fellow in the Chinese
University of Hong Kong. The work is supported in part by the NSF of China.}

\begin{description}
\item  \newpage\ {\bf References}

\item[1]  L. H. Yu and C. P. Sun, {\it Phys. Rev}., {\bf A49}(1994)592.

\item[2]  C. P.\ Sun and L. H. Yu, {\it Phys. Rev}., {\bf A51}(1995)1845.

\item[3]  A. O. Caldeira and A. J. Leggett, {\it Ann. Phys}. (N.Y.), {\bf 149%
}(1983)374.

\item[4]  M. Lax, {\it J. Phys. Chem. Solid,} {\bf 25}(1964)487; {\it Phys.
Rev}., {\bf A145}(1966)110.

\item[5]  P. Ullersma, {\it Physica}, {\bf 32}(1966)27.

\item[6]  G. W.Ford, J. T. Lewis and R. F. O'connell, {\it Phys. Rev}., {\bf %
A37}(1988)4419.

\item[7]  E. Kanai, {\it Prog. Theor. Phys}., {\bf 3}(1948)440.

\item[8]  P. Calderora, {\it Nuovo Cimento}, {\bf 18}(1941)393.

\item[9]  W. H. Louisell{\bf ,{\it \ Quantum Statistical Properties of
Radiation}, }John Wiley \& Sons, NewYork, 1973.
\end{description}

\end{document}